# A New Metallicity Calibration for Dwarfs for the RGU−Photometry

Salih KARAALİ, Selçuk BİLİR

*Istanbul University Science Faculty Department of Astronomy and Space Sciences,
34452 Istanbul - TURKEY*

## Abstract

We adopted the procedure of Carney to obtain a metallicity calibration for dwarfs for the RGU−photometry. For this purpose we selected 76 dwarfs of different metallicities from Carney, and Strobel et al., and evaluated their $\delta(U-G)$ ultra−violet excess relative to Hyades by transforming their UBV magnitudes to RGU via metallicity dependent equations of Ak−Güngör. The $\delta_{0.6}/\delta M$ normalized factors of Sandage transform $\delta(U-G)$ excess at any G−R to $\delta \equiv \delta_{1.08}$, i.e.: the ultra−violet excess at G−R = 1.08 mag, corresponding to B−V = 0.60 mag in the UBV−system. Finally, the ($\delta$, [Fe/H]) couples were fitted by the equation [Fe/H] = 0.11 − 2.22 $\delta$ − 7.95 $\delta^2$. This calibration covers the metallicity interval (−2.20, +0.20) dex.

## 1. Introduction

Metallicity plays an important role in the Galactic structure. Although mean metal−abundances were attributed to three main Galactic components, i.e.: Population I (Thin Disk), Intermediate Population II (Thick Disk), and Extreme Population II (Halo) (cf. Norris [1]), recent works show that the metallicity distributions for these populations may well be multimodal (Norris [1], Carney [2], Karaali et al. [3]). More important is the metallicity gradient cited either for populations individually or for a region of the Galaxy. Examples can be found in Reid and Majewski [4], and Chiba and Yoshii [5]. The importance is concerned with the formation of the Galaxy, as explained in the following: The existence of a metallicity gradient for any component of the Galaxy means that it formed by dissipative collapse. The pioneers of this suggestion are Eggen, Lynden−Bell, and Sandage [6, ELS]. A discussion of the current status of this model is provided by Gilmore, Wyse, and Kuijken [7]. Many later analyses followed (e.g. Yoshii and Saio [8], Norris, Bessel, and Pickles [9, NBP], Norris [10], Sandage and Fouts [11], Carney, Latham, and Laird [12], Norris and Ryan [13], Beers and Sommer−Larsen [14]). From these works, an alternative picture has emerged, suggesting that the collapse of the Galaxy occured slowly. This picture was postulated largely on a supposed wide age range in the globular clustar system (Searle and Zinn [15, SZ], Schuster and Nissen [16]). SZ especially argued that the Galactic halo was not formed in an order of collapse, but from the merger or accretion of numerous fragments, such as dwarf−type galaxies. Such a scenario indicates no metallicity gradient or younger and even more metal−rich objects at the outermost part of the Galaxy. The globular cluster age range supposition has been disproved by recent analyses (Rosenberg et al. [17]), while the number of young field halo stars has been shown to be extremely small, inconsistent with this model, by Unavane, Wyse, and Gilmore [18], Preston and Sneden [19], Gilmore [20].

A clear metallicity gradient is highly dependent on the precise metallicity determination. In the RGU photometry, the figures of Buser and Fenkart [21] and Buser et al. [22], calibrated as a function of both G−R colour index and the ultra−violet excess (relative to zero iso−metallicity line) is used. Whereas, the metallicity equation of Carney [23], i.e.: [Fe/H] = 0.11 − 2.90 δ − 18.68 δ$^2$, is prefered in the UBV photometry. The large sample of Carney includes stars of different categories, such as dwarfs, subgiants, and close binaries, and the calibration for this relationship extends down to [Fe/H] = −2.45 dex, however, it is assumed to be valid for [Fe/H] ≥ −1.75 dex (Laird, Carney, and Latham [24, LCL], Gilmore, Wyse, and Jones [25]). We thought to contribute this topic by obtaining a metallicity calibration for dwarfs for the RGU photometry, similar to that of Carney [23], and that is the main goal of this paper.

## 2. The Method

The method consists of adopting the procedure of Carney [23] to the RGU photometry. Two steps were followed for our purpose: in the first step, the UBV data for 52 and 24 dwarfs taken from Carney [23] and Strobel et al. [26], respectively, are transformed to the RGU data by means of metallicity dependent equations of Ak−Güngör [27], and the (U−G, G−R) main−sequence of Hyades, transformed from UBV to RGU by the same formulae, is used as a standard sequence for ultra−violet excess evaluation. The transformation formulae just cited or those of Buser [28] may be used to show that the factors given by Sandage [29] for the UBV photometry may also be used for the normalization of these excesses, as explained in the following:

The equations which transform the U−B and B−V colour indices of a star to the G−R and U−G colour indices are in the form of,

$$G-R = a_1(U-B) + b_1(B-V) + c_1 \quad (1)$$

$$U-G = a_2(U-B) + b_2(B-V) + c_2 \quad (2)$$

where $a_i$, $b_i$, and $c_i$ (i = 1, 2) are parameters to be determined. Let us write equation (2) for two stars with the same B−V (or equivalently G−R), i.e.: for a Hyades star (H) and for a star (*) whose ultra−violet excess would be normilized,

$$(U-G)_H = a_2(U-B)_H + b_2(B-V) + c_2 \quad (3)$$

$$(U-G)_* = a_2(U-B)_* + b_2(B-V) + c_2 \quad (4)$$

Then, the ultra−violet excess for the star in question, relative to the Hyades star is,

$$(U-G)_H - (U-G)_* = a_2[(U-B)_H - (U-B)_*] \quad (5)$$

or with the accustomed notation,

$$\delta(U-G) = a_2 \, \delta(U-B) \quad (6)$$

Now, for another star with the same metal−abundance [Fe/H] but with B−V = 0.6 mag, (or its equivalence G−R = 1.08 mag) we get, in the same way,

$$\delta(U-G)_{1.08} = a_2 \, \delta(U-B)_{0.6} \quad (7)$$

Equations (6) and (7) gives,

$$\frac{\delta(U-G)_{1.08}}{\delta(U-G)} = \frac{\delta(U-B)_{0.6}}{\delta(U-B)} = f \quad (8)$$

Where $f$ is the factor in question. Hence, $\delta(U-G)$ excess in the RGU photometry can be normilesed by the same $f$ factors used in UBV photometry. Table 1 gives the normalization factors taken from Sandage [29]. [Fe/H] metallicities and UBV data from Carney [23] and Strobel et al. [26], and their corresponding RGU data are given in Table 2a and Table 2b, respectively. Dwarfs in Table 2a are identified according to their spectral types, whereas dwarfs in Table 2b have large surface gravities, i.e.: log $g$ > 4.0.

**Table 1.** Hyades main-sequence and the normalization factor (*f*) of Sandage.

| (B-V)$_o$ | (U-B)$_o$ | *f* | (B-V)$_o$ | (U-B)$_o$ | *f* | (B-V)$_o$ | (U-B)$_o$ | *f* |
|---|---|---|---|---|---|---|---|---|
| 0.35 | 0.03 | 1.24 | 0.66 | 0.20 | 1.04 | 0.97 | 0.78 | 1.71 |
| 0.36 | 0.03 | 1.23 | 0.67 | 0.22 | 1.06 | 0.98 | 0.80 | 1.74 |
| 0.37 | 0.02 | 1.22 | 0.68 | 0.23 | 1.07 | 0.99 | 0.82 | 1.78 |
| 0.38 | 0.02 | 1.21 | 0.69 | 0.25 | 1.09 | 1.00 | 0.84 | 1.82 |
| 0.39 | 0.01 | 1.20 | 0.70 | 0.25 | 1.10 | 1.01 | 0.86 | 1.87 |
| 0.40 | 0.01 | 1.19 | 0.71 | 0.27 | 1.12 | 1.02 | 0.88 | 1.92 |
| 0.41 | 0.01 | 1.18 | 0.72 | 0.29 | 1.14 | 1.03 | 0.90 | 1.96 |
| 0.42 | 0.01 | 1.17 | 0.73 | 0.30 | 1.15 | 1.04 | 0.93 | 2.01 |
| 0.43 | 0.00 | 1.17 | 0.74 | 0.33 | 1.17 | 1.05 | 0.94 | 2.06 |
| 0.44 | 0.00 | 1.16 | 0.75 | 0.34 | 1.19 | 1.06 | 0.96 | 2.16 |
| 0.45 | 0.00 | 1.15 | 0.76 | 0.36 | 1.21 | 1.07 | 0.97 | 2.27 |
| 0.46 | 0.01 | 1.14 | 0.77 | 0.38 | 1.23 | 1.08 | 0.98 | 2.37 |
| 0.47 | 0.01 | 1.13 | 0.78 | 0.40 | 1.25 | 1.09 | 0.99 | 2.48 |
| 0.48 | 0.02 | 1.13 | 0.79 | 0.42 | 1.27 | 1.10 | 1.01 | 2.58 |
| 0.49 | 0.03 | 1.12 | 0.80 | 0.43 | 1.29 | 1.11 | 1.03 | 2.58 |

**Table 1. (cont.)**

| (B-V)$_o$ | (U-B)$_o$ | $f$ | (B-V)$_o$ | (U-B)$_o$ | $f$ | (B-V)$_o$ | (U-B)$_o$ | $f$ |
|---|---|---|---|---|---|---|---|---|
| 0.50 | 0.03 | 1.11 | 0.81 | 0.45 | 1.31 | 1.12 | 1.05 | 2.58 |
| 0.51 | 0.04 | 1.09 | 0.82 | 0.47 | 1.34 | 1.13 | 1.07 | 2.58 |
| 0.52 | 0.05 | 1.08 | 0.83 | 0.49 | 1.36 | 1.14 | 1.08 | 2.82 |
| 0.53 | 0.06 | 1.06 | 0.84 | 0.52 | 1.39 | 1.15 | 1.10 | 2.82 |
| 0.54 | 0.07 | 1.05 | 0.85 | 0.54 | 1.41 | 1.16 | 1.12 | 2.82 |
| 0.55 | 0.08 | 1.03 | 0.86 | 0.56 | 1.44 | 1.17 | 1.14 | 2.82 |
| 0.56 | 0.09 | 1.02 | 0.87 | 0.58 | 1.47 | 1.18 | 1.15 | 3.10 |
| 0.57 | 0.10 | 1.02 | 0.88 | 0.60 | 1.49 | 1.19 | 1.16 | 3.44 |
| 0.58 | 0.11 | 1.01 | 0.89 | 0.62 | 1.52 | 1.20 | 1.17 | 3.44 |
| 0.59 | 0.12 | 1.01 | 0.90 | 0.64 | 1.55 | 1.21 | 1.18 | 4.43 |
| 0.60 | 0.13 | 1.00 | 0.91 | 0.66 | 1.57 | 1.22 | 1.19 | 4.77 |
| 0.61 | 0.14 | 1.01 | 0.92 | 0.68 | 1.58 | 1.23 | 1.19 | 6.20 |
| 0.62 | 0.16 | 1.01 | 0.93 | 0.70 | 1.60 | 1.24 | 1.20 | 7.75 |
| 0.63 | 0.17 | 1.02 | 0.94 | 0.71 | 1.61 | 1.25 | 1.21 | 7.75 |
| 0.64 | 0.18 | 1.02 | 0.95 | 0.74 | 1.63 | | | |
| 0.65 | 0.19 | 1.03 | 0.96 | 0.76 | 1.67 | | | |

**Table 2a.** [Fe/H] metallicities and UBV data from Carney and their corresponding RGU data. The metallicities in columns 3 and 7 correspond to the adopted values from Carney, and the computed ones by means of equation [Fe/H] = 0.11 − 2.22 $\delta_{(1.08)}$ − 7.95 $\delta^2_{(1.08)}$; $\delta_{(1.08)}$ in column 6 is the ultra–violet excess relative to Hyades, reduced to the colour index G−R = 1.08, and finally column 8 includes the differences of metallicities in columns 3 and 7.

| 1 | 2 | 3 | 4 | 5 | 6 | 7 | 8 |
|---|---|---|---|---|---|---|---|
| B–V | U–B | [Fe/H]$_{ad}$ | G–R | U–G | $\delta_{(1.08)}$ | [Fe/H]$_c$ | Δ[Fe/H] |
| 0.62 | 0.11 | 0.12 | 1.11 | 1.46 | 0.07 | -0.10 | 0.22 |
| 0.72 | 0.21 | -0.36 | 1.22 | 1.57 | 0.13 | -0.34 | -0.02 |
| 0.42 | -0.07 | -0.40 | 0.87 | 1.20 | 0.12 | -0.30 | -0.10 |
| 0.61 | 0.02 | -0.50 | 1.09 | 1.33 | 0.17 | -0.53 | 0.03 |
| 0.49 | 0.00 | -0.10 | 0.95 | 1.31 | 0.04 | 0.00 | -0.10 |
| 0.44 | -0.12 | -0.35 | 0.90 | 1.15 | 0.18 | -0.58 | 0.23 |
| 0.46 | -0.24 | -1.92 | 0.96 | 1.01 | 0.38 | -1.86 | -0.06 |
| 0.68 | 0.19 | 0.10 | 1.18 | 1.56 | 0.07 | -0.10 | 0.20 |
| 0.64 | 0.07 | -0.37 | 1.13 | 1.40 | 0.14 | -0.39 | 0.02 |
| 0.60 | -0.02 | -0.34 | 1.08 | 1.29 | 0.19 | -0.63 | 0.29 |

**Table 2a. (cont.)**

| 1 | 2 | 3 | 4 | 5 | 6 | 7 | 8 |
|---|---|---|---|---|---|---|---|
| B–V | U–B | [Fe/H]$_{ad}$ | G–R | U–G | $\delta_{(1.08)}$ | [Fe/H]$_c$ | $\Delta$[Fe/H] |
| 0.71 | 0.22 | -0.34 | 1.21 | 1.58 | 0.09 | -0.18 | -0.16 |
| 0.54 | -0.08 | -0.57 | 1.01 | 1.21 | 0.21 | -0.73 | 0.16 |
| 0.86 | 0.38 | -1.33 | 1.39 | 1.77 | 0.36 | -1.70 | 0.37 |
| 0.59 | 0.02 | -0.32 | 1.07 | 1.34 | 0.14 | -0.39 | 0.07 |
| 0.46 | -0.01 | 0.17 | 0.91 | 1.30 | 0.02 | 0.06 | 0.11 |
| 0.59 | 0.08 | 0.06 | 1.07 | 1.42 | 0.06 | -0.07 | 0.13 |
| 0.57 | -0.07 | -0.80 | 1.06 | 1.22 | 0.25 | -0.96 | 0.16 |
| 0.61 | -0.12 | -1.73 | 1.12 | 1.16 | 0.38 | -1.86 | 0.13 |
| 0.70 | 0.18 | -0.62 | 1.20 | 1.53 | 0.12 | -0.30 | -0.32 |
| 0.47 | -0.06 | -0.52 | 0.93 | 1.22 | 0.14 | -0.39 | -0.13 |
| 0.62 | 0.07 | -0.27 | 1.10 | 1.39 | 0.12 | -0.30 | 0.03 |
| 0.43 | -0.23 | -2.05 | 0.94 | 1.02 | 0.36 | -1.70 | -0.35 |
| 0.62 | 0.08 | -0.04 | 1.11 | 1.43 | 0.10 | -0.22 | 0.18 |
| 0.60 | 0.05 | -0.23 | 1.09 | 1.39 | 0.11 | -0.26 | 0.03 |

**Table 2a. (cont.)**

| 1 | 2 | 3 | 4 | 5 | 6 | 7 | 8 |
|---|---|---|---|---|---|---|---|
| B–V | U–B | [Fe/H]$_{ad}$ | G–R | U–G | $\delta_{(1.08)}$ | [Fe/H]$_c$ | $\Delta$[Fe/H] |
| 0.74 | 0.23 | -0.27 | 1.24 | 1.59 | 0.13 | -0.34 | 0.07 |
| 0.44 | -0.13 | -0.72 | 0.90 | 1.14 | 0.19 | -0.63 | -0.09 |
| 0.43 | -0.22 | -1.70 | 0.94 | 1.03 | 0.35 | -1.63 | -0.07 |
| 0.59 | 0.05 | -0.04 | 1.07 | 1.39 | 0.09 | -0.18 | 0.14 |
| 0.55 | -0.03 | -0.31 | 1.02 | 1.27 | 0.15 | -0.43 | 0.12 |
| 0.58 | 0.08 | 0.16 | 1.06 | 1.42 | 0.05 | -0.04 | 0.20 |
| 0.52 | -0.08 | -0.64 | 0.99 | 1.21 | 0.19 | -0.63 | -0.01 |
| 0.77 | 0.29 | -0.13 | 1.29 | 1.69 | 0.16 | -0.48 | 0.35 |
| 0.79 | 0.33 | -0.55 | 1.30 | 1.71 | 0.16 | -0.48 | -0.07 |
| 0.64 | 0.06 | -0.49 | 1.13 | 1.38 | 0.16 | -0.48 | -0.01 |
| 0.60 | 0.10 | 0.03 | 1.08 | 1.45 | 0.03 | 0.03 | 0.00 |
| 0.48 | -0.03 | -0.19 | 0.94 | 1.28 | 0.07 | -0.10 | -0.09 |
| 0.46 | -0.20 | -1.67 | 0.97 | 1.06 | 0.33 | -1.49 | -0.18 |
| 0.53 | -0.07 | -0.63 | 1.00 | 1.22 | 0.19 | -0.63 | 0.00 |

**Table 2a. (cont.)**

| 1 | 2 | 3 | 4 | 5 | 6 | 7 | 8 |
|---|---|---|---|---|---|---|---|
| B–V | U–B | [Fe/H]$_{ad}$ | G–R | U–G | $\delta_{(1.08)}$ | [Fe/H]$_c$ | $\Delta$[Fe/H] |
| 0.65 | 0.08 | -0.38 | 1.14 | 1.41 | 0.15 | -0.43 | 0.05 |
| 0.56 | -0.01 | -0.54 | 1.03 | 1.29 | 0.14 | -0.39 | -0.15 |
| 0.62 | 0.06 | -0.47 | 1.10 | 1.38 | 0.13 | -0.34 | -0.13 |
| 0.48 | -0.10 | -0.94 | 0.98 | 1.18 | 0.22 | -0.79 | -0.15 |
| 0.58 | -0.03 | -0.69 | 1.06 | 1.27 | 0.20 | -0.68 | -0.01 |
| 0.52 | -0.08 | -0.46 | 0.99 | 1.21 | 0.19 | -0.63 | 0.17 |
| 0.81 | 0.34 | -0.87 | 1.34 | 1.72 | 0.25 | -0.96 | 0.09 |
| 0.61 | -0.15 | -1.80 | 1.12 | 1.12 | 0.42 | -2.18 | 0.38 |
| 0.53 | -0.06 | -1.10 | 1.03 | 1.23 | 0.20 | -0.68 | -0.42 |
| 0.51 | -0.16 | -1.42 | 0.99 | 1.11 | 0.30 | -1.28 | -0.14 |
| 0.48 | -0.13 | -0.68 | 0.94 | 1.15 | 0.21 | -0.73 | 0.05 |
| 0.41 | -0.15 | -0.53 | 0.86 | 1.11 | 0.23 | -0.85 | 0.32 |
| 0.50 | -0.13 | -0.51 | 0.97 | 1.15 | 0.23 | -0.85 | 0.34 |
| 0.67 | 0.20 | 0.12 | 1.17 | 1.57 | 0.04 | 0.00 | 0.12 |

**Table 2b.** [Fe/H] metallicities and UBV data from Cayrel de Strobel et al. and their corresponding RGU data. Symbols as in Table 2a.

| 1 | 2 | 3 | 4 | 5 | 6 | 7 | 8 |
|---|---|---|---|---|---|---|---|
| B–V | U–B | [Fe/H]$_{ad}$ | G–R | U–G | $\delta_{(1.08)}$ | [Fe/H]$_c$ | $\Delta$[Fe/H] |
| 0.30 | -0.31 | -2.15 | 0.80 | 0.92 | 0.48 | -2.70 | 0.55 |
| 0.64 | -0.11 | -2.11 | 1.15 | 1.17 | 0.42 | -2.18 | 0.07 |
| 0.39 | -0.27 | -2.06 | 0.89 | 0.97 | 0.39 | -1.93 | -0.13 |
| 0.41 | -0.28 | -2.04 | 0.91 | 0.96 | 0.40 | -2.01 | -0.03 |
| 0.43 | -0.27 | -2.01 | 0.93 | 0.97 | 0.42 | -2.18 | 0.17 |
| 0.26 | -0.28 | -2.00 | 0.77 | 0.95 | 0.47 | -2.61 | 0.61 |
| 0.61 | -0.13 | -2.00 | 1.12 | 1.15 | 0.39 | -1.93 | -0.07 |
| 0.61 | -0.15 | -1.86 | 1.12 | 1.12 | 0.42 | -2.18 | 0.32 |
| 0.56 | -0.12 | -1.60 | 1.07 | 1.16 | 0.32 | -1.42 | -0.18 |
| 0.59 | -0.13 | -1.60 | 1.00 | 1.15 | 0.36 | -1.70 | 0.10 |
| 0.53 | -0.09 | -1.60 | 1.05 | 1.19 | 0.27 | -1.09 | -0.51 |
| 0.35 | -0.24 | -1.30 | 0.84 | 1.01 | 0.35 | -1.63 | 0.33 |
| 0.52 | -0.14 | -1.22 | 1.01 | 1.14 | 0.28 | -1.15 | -0.07 |

**Table 2b. (cont.)**

| 1 | 2 | 3 | 4 | 5 | 6 | 7 | 8 |
|---|---|---|---|---|---|---|---|
| B–V | U–B | [Fe/H]$_{ad}$ | G–R | U–G | $\delta_{(1.08)}$ | [Fe/H]$_c$ | $\Delta$[Fe/H] |
| 0.55 | -0.06 | -1.07 | 1.05 | 1.23 | 0.23 | -0.85 | -0.22 |
| 0.54 | -0.12 | -0.99 | 1.03 | 1.16 | 0.28 | -1.15 | 0.16 |
| 0.47 | -0.14 | -0.99 | 0.96 | 1.14 | 0.24 | -0.90 | -0.09 |
| 0.45 | -0.12 | -0.84 | 0.95 | 1.16 | 0.21 | -0.73 | -0.11 |
| 0.69 | 0.10 | -0.71 | 1.18 | 1.44 | 0.20 | -0.68 | -0.03 |
| 0.69 | 0.17 | -0.60 | 1.18 | 1.52 | 0.11 | -0.26 | -0.34 |
| 0.77 | 0.38 | -0.23 | 1.29 | 1.80 | 0.03 | 0.03 | -0.26 |
| 0.93 | 0.69 | -0.06 | 1.48 | 2.19 | 0.03 | 0.03 | -0.09 |
| 0.82 | 0.50 | 0.01 | 1.35 | 1.95 | -0.01 | 0.14 | -0.13 |
| 0.57 | 0.04 | 0.06 | 1.05 | 1.37 | 0.09 | -0.18 | 0.24 |
| 0.85 | 0.49 | 0.09 | 1.39 | 1.94 | 0.12 | -0.30 | 0.39 |

**Table 3.** Metallicity distribution for 76 stars taken from Carney, and Cayrel de Strobel et al. Mean metallicities (column 3) and mean ultra−violet excesses relative to Hyades, reduced to the colour−index G−R = 1.08 (column 4) are also given. Column 5 gives the number of stars for each metallicity interval.

| 1 | 2 | 3 | 4 | 5 |
|---|---|---|---|---|
| [Fe/H] | | <[Fe/H]> | <$\delta_{1.08}$> | N |
| -2.20 | -2.00 | -2.10 | 0.42 | 8 |
| -2.00 | -1.70 | -1.85 | 0.39 | 5 |
| -1.70 | -1.40 | -1.55 | 0.32 | 5 |
| -1.40 | -1.10 | -1.25 | 0.30 | 4 |
| -1.10 | -0.90 | -1.00 | 0.24 | 4 |
| -0.90 | -0.70 | -0.80 | 0.22 | 5 |
| -0.70 | -0.60 | -0.65 | 0.17 | 6 |
| -0.60 | -0.50 | -0.55 | 0.18 | 7 |
| -0.50 | -0.40 | -0.45 | 0.15 | 4 |
| -0.40 | -0.30 | -0.35 | 0.14 | 8 |
| -0.30 | -0.20 | -0.25 | 0.10 | 4 |
| -0.20 | 0.00 | -0.10 | 0.08 | 6 |
| 0.00 | 0.10 | 0.05 | 0.06 | 6 |
| 0.10 | 0.20 | 0.15 | 0.05 | 4 |

In the second step, 76 stars given in Tables 2a and 2b are separated into 14 metallicity intervals, with different scales, such as to involve almost equal stars in number, and the least–square method is used to obtain a calibration between the normilazied ultra–violet excess $\delta_{1.08}$ and metallicity [Fe/H] (Table 3). Such a separation provides equal–weight for 14 points in Fig. 1 which represent the mean metallicities and mean $\delta_{1.08}$ excesses in each metallicity interval in Table 3. The constant term $a_o$ in the equation,

$$[Fe/H] = a_o + a_1 \delta_{1.08} + a_2 \delta^2_{1.08} \qquad (9)$$

is assumed to be $a_o = 0.11$ for the consistency with the metallicity of Hyades, cited by Carney [23]. The least–square method gives $a_1 = -2.22$ and $a_2 = -7.95$ thus,

$$[Fe/H] = 0.11 - 2.22\delta_{1.08} - 7.95\, \delta^2_{1.08} \qquad (10)$$

The differences between the metallicities evaluated by means of equation (10) and the original ones, ie.: $\Delta$[Fe/H], (column 8, in Tables 2a and 2b) versus the original metallicities is given in Fig. 2. The differences are large only for a few metal–poor stars and the scatter relative to the line $\Delta$[Fe/H] = 0.0 dex is small. Actually the mean of the differences (for all stars) is 0.02 dex and the probable error for the mean, p.e. = ± 0.15 dex indicating that the new calibration would be used with a good accuracy.

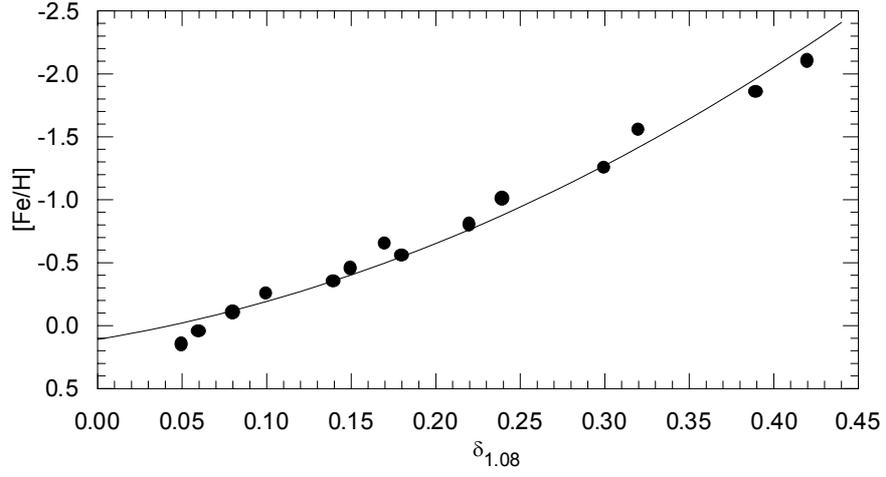

**Fig. 1.** [Fe/H] metallicity versus normilized $\delta_{1.08}$ ultra–violet excess for the RGU photometry.

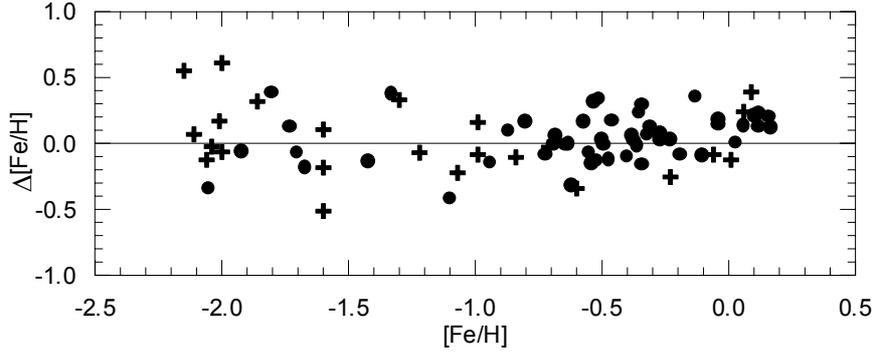

**Fig. 2.** $\Delta$[Fe/H] versus metallicity, where $\Delta$[Fe/H] is the difference between the original metallicities and the evaluated ones, by means of the new calibration, [Fe/H] = 0.11 − 2.22 $\delta_{1.08}$ − 7.95 $\delta^2_{1.08}$. Symbols: (•) stars from Table 2a, and (+) stars from Table 2b.

## 3. Dicussion

The procedure used by Buser and Fenkart [21] for the metallicity determination for a dwarf star is based on the combination of its ultra−violet excesses relative to Population I (zero metallicity) dwarfs and its G−R colour index, in a diagram which requires considerable large work, whereas the calibration given in our work is more practical. Although the procedure in the work of Buser and Fenkart [21] provides a larger metallicity interval relative to ours, i.e.: down to [Fe/H] = −3.0 dex and [Fe/H] = −2.2 dex respectively, the number of stars with [Fe/H] < −2.2 dex in our Galaxy are rary, thus the new calibration does not bring any considerable restriction in the metallicity works cocerned with the Galactic structure.

The comparison of the new calibration with the one of Carney [23] shows that our calibration has some adventages, probably due to additional metal−poor stars, as explained in the following: **(1)** the new calibration is available for stars with [Fe/H] ≥ −2.20 dex, whereas the other calibration is limited with [Fe/H] ≥ −1.75 dex. **(2)** the gradient for the diagram [Fe/H] versus normalized UV−excess, $\delta_{1.08}$, in this work is steeper than the corresponding one in the work of Carney [23] (although the figure of Carney is not given here, this can be deduced by the comparison of the scales in uv-excess in two works) providing more accurate metallicities. **(3)** the diagram Δ[Fe/H] versus original metallicities in our work (Fig. 2) is better than the one for the data of Carney (Fig. 3), where Δ[Fe/H] is the difference between the original metallicities and the evaluated ones by the corresponding equation. Actually, the mean and standard deviation for Δ[Fe/H], for all stars, in our work are <[Fe/H]> = 0.02 dex, and $s$ = ± 0.22 dex, respectively, whereas they are <[Fe/H]> = 0.03 dex, and *s* = ± 0.27 dex for the data evaluated via the formula of Carney, for stars only with Δ[Fe/H] < +1.00 dex (four stars with Δ[Fe/H] ≥ + 1.00 dex in Fig. 3 are not included for these evaluations). Hence, we hope that we would contribute to the metallicity researches by the new calibration.

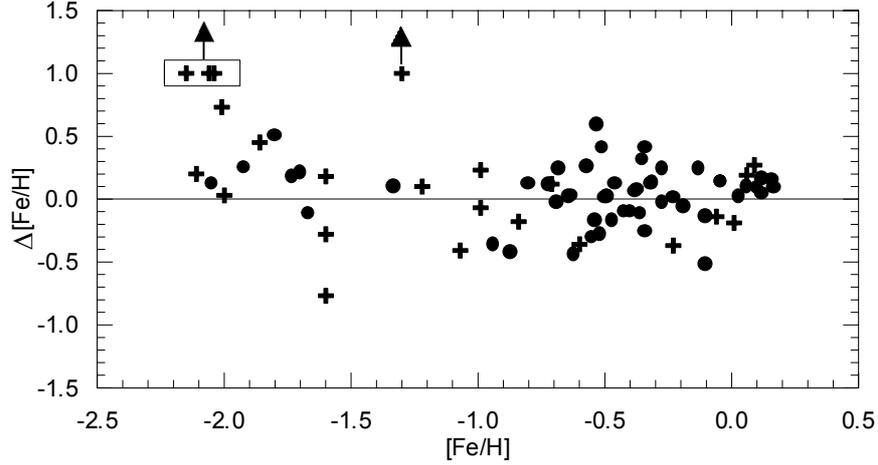

**Fig. 3.** Δ[Fe/H] versus metallicity, where Δ[Fe/H] is the difference between the original metallicities and the evaluated ones by means of the Carney's equation, i.e.: [Fe/H] = 0.11 − 2.90 $\delta_{0.6}$ − 18.68 $\delta^2_{0.6}$. Symbols as in Fig. 2.